\documentclass{elsart4-1}

\usepackage{graphicx}

\usepackage{amsmath,amssymb}

\usepackage[english,francais]{babel}

\newtheorem{e-proposition}[theorem]{Proposition}

\newtheorem{e-definition}[theorem]{Definition\rm}

\renewcommand{\theequation}{\arabic{equation}}
\def\og{\leavevmode\raise.3ex\hbox{$\scriptscriptstyle\langle\!\langle$~}}
\def\fg{\leavevmode\raise.3ex\hbox{~$\!\scriptscriptstyle\,\rangle\!\rangle$}}

\DeclareMathOperator{\Tr}{Tr}

\begin{document}


\centerline{Physics or Astrophysics/Header}
\begin{frontmatter}


\selectlanguage{english}

\title{Nonlinear phenomena in quantum thermoelectrics and heat}


\selectlanguage{english}
\author[authorlabel1]{David S\'anchez},
\ead{david.sanchez@uib.es}
\author[authorlabel2]{Rosa L\'opez}
\ead{rosa.lopez-gonzalo@uib.es}

\address[authorlabel1]{Instituto de F\'{\i}sica Interdisciplinar y Sistemas Complejos (UIB-CSIC), Campus Universitat Illes Balears, E-07122 Palma de Mallorca, Spain}
\address[authorlabel2]{Instituto de F\'{\i}sica Interdisciplinar y Sistemas Complejos (UIB-CSIC), Campus Universitat Illes Balears, E-07122 Palma de Mallorca, Spain}


\medskip
\begin{center}
{\small }
\end{center}

\begin{abstract}
We review recent developments in nonlinear quantum transport through nanostructures and mesoscopic systems
driven by thermal gradients or in combination with voltage biases. Low-dimensional conductors are excellent platforms
to analyze both the thermoelectric and heat dynamics beyond linear response because due to their small
size a small temperature difference applied across regions gives rise to large thermal biases.
We offer a theoretical discussion based on the scattering approach to highlight the differences between
the linear and the nonlinear regimes of transport. We discuss recent experiments on quantum dots and
molecular junctions subjected to strong temperature differences. Theoretical predictions concerning
the Kondo effect and heat rectification proposals are briefly examined. An important issue is the
calculation of thermoelectric efficiencies including nonlinearities. Cross Seebeck effects and nonlinear
spin filtering arise in superconductors and topological insulators while mixed noises
between charge and heat currents are also considered. Finally, we provide an outlook
on the possible future directions of the field.

\vskip 0.5\baselineskip


 \vskip 0.5\baselineskip}
\end{abstract}
\end{frontmatter}


\selectlanguage{english}

\section{Introduction}

Nonlinear phenomena are common in nature. Certain effects occurring in the electronic
transport through solids (diode behavior~\cite{sze},
negative differential resistance in tunnel junctions~\cite{esaki}, Gunn oscillations~\cite{gunn})
are unique to the nonlinear regime and can be only explained within
theoretical models that include a nonlinear relation between the forces and the fluxes.
In the quantum realm, electronic transport is dictated by the scattering properties of the conductor. The latter
is typically attached to multiple reservoirs with well defined electrochemical potentials and temperatures deep inside
the reservoirs. As a consequence of voltage or thermal shifts applied to these terminals, currents associated to both
charge and energy flow through the sample. Importantly, this is a nonequilibrium problem and therefore
nonequilibrium responses must be, in general, determined. In the linear regime (first-order perturbation in the shifts),
an expansion around the equilibrium point allows us to express the responses in terms of the equilibrium
potential. Linear responses then obey symmetry and reciprocity relations, as demonstrated by Onsager~\cite{onsager}.
However, beyond linear
response one needs to find the charge buildup on the conductor induced by the external shifts
which in turn affects the sample potential. The problem thus becomes self-consistent and is generally difficult
to solve. This paper presents an account of the most recent advances toward a clear understanding
of the nonlinear transport properties of mesoscopic and nanoscopic systems when both driving forces
(voltages and temperature biases) are active at the same time. Accordingly, it is expected that transport
is dominated by thermoelectric effects and that not only charge but also heat dynamics
play a significant role. The subject is important due to the nanostructures' ability to generate
large gradients with moderate applied shifts
(e.g., superlattices~\cite{ven01} and molecular junctions~\cite{red07})
and to their impact on the quest for energy harvesters and
thermoelectric coolers with improved efficiencies~\cite{vin10,dub11,here13,san13b,zim16}.

The Seebeck effect is the generation of a voltage across the conductor that counterbalances the applied
temperature difference in open circuit conditions (zero net current)~\cite{book}. In the linear regime, the Seebeck coefficient
or thermopower
is thus given by the ratio of two responses: the thermoelectric and the electric responses. For macroscopic conductors
the thermal gradient is small and the thermopower is independent of the temperature and voltage biases.
In turn, the Peltier effect is the creation of a heat flow in response to an applied voltage in isothermal conditions
(zero thermal gradient). In contrast to the Joule heat, which is quadratic in voltage, the Peltier heat is reversible:
it changes direction if the electric field is reversed. Due to reciprocity, both Seebeck and Peltier coefficients are nicely
related to each other. Whether or not this symmetry relation holds beyond linear response is one of the issues
to be tackled in a nonlinear thermoelectric transport formalism. Finally, it is worth noting that heat can be also
generated with a thermal difference in the isoelectric case (no applied voltage). Up to first order in the temperature
bias, the response is constant and given by the thermal conductance. This Fourier law is not generally met in
the nonlinear regime, where the heat now becomes a function of higher orders in a temperature expansion,
and thermal rectification effects may arise~\cite{ter02}.
This is an important point since the thermoelectric figure of merit, which is typically
written in terms of linear responses only
(electric conductance, thermopower and thermal conductance), should be generalized to the nonlinear regime
as a ratio between the generated power (nonlinear current times voltage) and the nonlinear heat flow.
Although this obviously has many practical implications, the interest in the scientific literature
has been complemented with an elucidation of the fundamentals of electronic and heat quantum transport
subjected to large fields. The results obtained in the last few years demonstrate that mesoscopic conductors
behave in a remarkable way when they are driven far from equilibrium with both electric and thermal means.

We begin our discussion with a simple problem.
Consider a two-terminal phase-coherent conductor. (The case of an arbitrary number of leads will be treated
later). Let $I$ ($J$) denote the electric charge (heat) current measured at the left ($L$) contact. The scattering approach
yields the currents
\begin{align}
I&= \frac{2e}{h} \int dE \,\mathcal{T}(\mathcal{U},E) [f_L(E)-f_R(E)] \label{eq_I}\,,\\
J&= \frac{2}{h} \int dE\, (E-\mu_L) \mathcal{T}(\mathcal{U},E) [f_L(E)-f_R(E)]\label{eq_J}\,,
\end{align}
where spin degeneracy has been assumed. Here, $\mathcal{T}(\mathcal{U},E)$ is the transmission probability
for a carrier with energy $E$ traversing the conductor with a potential landscape given by the electrostatic potential
$\mathcal{U}=\mathcal{U}(V_L,V_R,\theta_L,\theta_R)$ that defines the screening properties of the conductor.
The potential is, quite generally, a function of position but here
we consider the homogeneous case for the sake of definiteness. Crucially, $\mathcal{U}$ contains
a nonequilibrium term that is a function
of the applied voltages, $V_L$ and $V_R$, and the temperature shifts, $\theta_L$ and $\theta_R$.
This term is determined from electron-electron interactions that govern screening effects out of equilibrium.
Hence, the transmission depends on the applied driving fields and the dependence will be strongest
for conductors with poor screening qualities as is usual in nanostructures. This dependence of the transmission
probability on voltage and temperature will primarily emerge
in the weakly nonlinear response (to leading order) because the Fermi distribution functions
in Eqs.~\eqref{eq_I} and~\eqref{eq_J} already depend on voltages and temperature shifts,
$f_\alpha(E)=1/[e^{(E-E_F-eV_\alpha)/(T+\theta_\alpha)}+1]$, where $E_F$ is the common Fermi energy
and $T$ is the background temperature. Therefore, it is natural to first examine the weakly nonlinear regime
since it will provide us with useful information on the nonequilibrium screening properties of the nanostructure.

\section{Scattering approach and weakly nonlinear transport}\label{sec-SA}

\begin{figure}[t]
\center
\includegraphics[angle=-0,width=0.5\textwidth,clip]{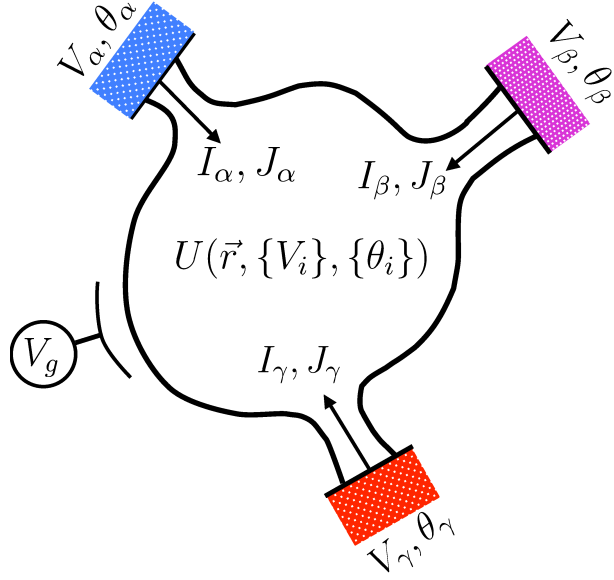}
\caption{(Color online.) A generic mesoscopic conductor is coupled to multiple terminals
labeled with $\alpha$, $\beta$, $\gamma$\ldots~These reservoirs have different voltage biases $V$
and temperature shifts $\theta$ away from a background temperature, causing
a flow of charge $I$ and heat $J$ that traverse the sample. Eventually, the conductor can be capacitively
coupled to a nearby gate with potential $V_g$. Importantly, the electrostatic landscape $\mathcal{U}$ of the nanostructure
depends in the nonequilibrium case on the set of voltage and thermal gradients applied to the coupled reservoirs.
}
\label{figsketch}
\end{figure}

A mesoscopic conductor is coupled to reservoirs labeled with Greek indices $\alpha$, $\beta$\ldots
(see Fig.~\ref{figsketch}).
These terminals are in local equilibrium with definite electrochemical potentials
$\mu_\alpha=E_F+eV_\alpha$ and temperatures $T_\alpha=T+\theta_\alpha$. The charge $I_\alpha$
and the heat $J_\alpha$ currents flowing into the conductor are respectively expressed as
\begin{align}\label{exactcurrents1}
I_\alpha &=\frac{2e}{h}\sum_\beta\int dE A_{\alpha\beta}(E,e\mathcal{U}) f_{\beta}(E)\,,\\
\label{exactcurrents2}
\mathcal{J}_\alpha&=\frac{2}{h}\sum_\beta\int dE (E-\mu_\alpha) A_{\alpha\beta}(E,e\mathcal{U}) f_{\beta}(E)\,,
\end{align}
where $A_{\alpha\beta}=\Tr [\delta_{\alpha\beta}-
s_{\alpha\beta}^\dagger s_{\alpha\beta}]$ and  $f_\beta(E)=1/(1+\exp{[(E-E_F-eV_\beta)/k_B T_\beta]})$.
The probability amplitude for a carrier to be scattered between leads $\beta$ and $\alpha$
is denoted with $s_{\alpha\beta}$. Here, we consider stationary fields only. Therefore, current conservation
demands that $\sum_\alpha I_\alpha=0$ while heat flows satisfy a conservation rule including the Joule
terms: $\sum_{\alpha} (J_{\alpha}+ I_{\alpha}V_\alpha)=0$. For the two terminal case, use of these
conservation laws leads to Eqs.~\eqref{eq_I} and~\eqref{eq_J} defining the transmission as
$\mathcal{T}={\rm Tr}\, s_{LR}^\dagger s_{LR}$.

Equations \eqref{exactcurrents1} and \eqref{exactcurrents2} express nonlinear relations between the fluxes,
$I_\alpha$ and $J_\alpha$, and the driving fields, $V_\alpha$ and $\theta_\alpha$. We can make an expansion
up to second order in the fields~\cite{san13,lop13},
\begin{align}\label{currents}
I_\alpha &=\sum_{\beta}G_{\alpha\beta}V_\beta
+\sum_{\beta}L_{\alpha\beta}\theta_\beta
+\sum_{\beta\gamma}G_{\alpha\beta\gamma}V_\beta V_\gamma 
+\sum_{\beta\gamma}L_{\alpha\beta\gamma}\theta_\beta \theta_\gamma
+2\sum_{\beta\gamma}M_{\alpha\beta\gamma}V_\beta \theta_\gamma\,,
\\ 
\label{eq_heatcurrent}\mathcal{J}_\alpha &=\sum_{\beta} R_{\alpha\beta}V_\beta
+\sum_{\beta}K_{\alpha\beta}\theta_\beta
+\sum_{\beta\gamma}R_{\alpha\beta\gamma}V_\beta V_\gamma 
+\sum_{\beta\gamma}K_{\alpha\beta\gamma}\theta_\beta \theta_\gamma
+2\sum_{\beta\gamma}H_{\alpha\beta\gamma}V_\beta \theta_\gamma\,,
\end{align}
where the linear responses are given by~\cite{siv86,but90}
\begin{subequations}\label{eq_linear}
\begin{align}
G_{\alpha\beta}&=\frac{2e^2}{h}\int dE A_{\alpha\beta}(E)\left[-\partial_E f(E)\right]\,,\\
L_{\alpha\beta}&=\frac{2e}{hT}\int dE (E-E_F) A_{\alpha\beta}(E)\left[-\partial_E f(E)\right]\,,\\
R_{\alpha\beta}&=\frac{2e}{h}\int dE (E-E_F) A_{\alpha\beta}(E)\left[-\partial_E f(E) \right]\,, \\
K_{\alpha\beta}&=\frac{2}{h}\int dE \frac{(E-E_F)^2}{T}A_{\alpha\beta}(E)\left[-\partial_E f(E) \right]\,.
\end{align}
\end{subequations}
It is important to remark that these four coefficients are evaluated at the equilibrium potential
(the electrostatic state characterized by $V_{\alpha}=0$ and $\theta_\alpha=0$ for all $\alpha$).
As a consequence, the basic principles of the theory of irreversible processes in thermodynamics can be invoked to
show that the thermoelectric (Seebeck) conductance $L$ and the electrothermal (Peltier) response $R$
are related via $R_{\alpha\beta}=TL_{\alpha\beta}$. There also exists an equivalence between
the electric $G$ and the thermal $K$ conductances but it only holds at very low temperatures
and for structureless contacts.
This can be seen from a leading-order Sommerfeld expansion of the latter equations,
\begin{subequations}\label{eq_sommerfeld}
\begin{align}
G_{\alpha\beta}&\simeq \frac{2e^2}{h} A_{\alpha\beta}(E_F) \,,\\
L_{\alpha\beta}&\simeq\frac{2e^2}{h} \frac{\pi^2 k_B^2T}{3}\partial_E A_{\alpha\beta}(E)|_{E=E_F} \,,\\
R_{\alpha\beta}&\simeq \frac{2e^2}{h} \frac{\pi^2 k_B^2T^2}{3}\partial_E A_{\alpha\beta}(E)|_{E=E_F} \,,\\
K_{\alpha\beta}&\simeq \frac{2e^2}{h} \frac{\pi^2 k_B^2T}{3e^2} A_{\alpha\beta}(E_F)\,,
\end{align}
\end{subequations}
from which it follows that $G_{\alpha\beta}/TK_{\alpha\beta}=L_0$ with $L_0\equiv \pi^2 k_B^2/3e^2$
the Lorenz number (Wiedemann-Franz law). We note that our discussion has assumed no magnetic effects.
In the presence of an external magnetic field $B$, the reciprocity relations must take into account
the opposite directions of the field. Microreversibility imposes that the equilibrium part of the electrostatic
potential be an even function of $B$, i.e., $\mathcal{U}_{\rm eq}(B)=\mathcal{U}_{\rm eq}(-B)$.
Hence, the matrices formed by the coefficients given by Eq.~\eqref{eq_linear}
satisfy $\mathbf{G}(B)=\mathbf{G}^t(-B)$, $\mathbf{K}(B)=\mathbf{K}^t(-B)$, and
$\mathbf{R}(B)=T\mathbf{L}^t(-B)$, with $t$ denoting transpose of the matrix.
The first equality is a cornerstone of mesoscopic physics~\cite{but86,ben86}.
In the absence of inelasticity (see, however, Refs.~\cite{san11,sai11,bra13}), one additionally obtains the relation
$\mathbf{L}(B)=\mathbf{L}^t(-B)$,
which has been experimentally demonstrated~\cite{mat14} using a ballistic cross junction
with an inserted asymmetric scatterer~\cite{mat12}.

Let us now turn to the nonlinear case. The main difference with the linear coefficients
is that the nonlinear responses depend on the voltage and temperature shifts.
This is clear from the explicit expressions:
\begin{subequations}\label{eq_nonlinear}
\begin{align}
G_{\alpha\beta\gamma}&=\frac{-e^2}{h}\int  dE 
\left(\frac{\partial A_{\alpha\beta}}{\partial V_\gamma } +\frac{\partial A_{\alpha\gamma}}{\partial V_\beta } + e\delta_{\beta\gamma} \partial_E A_{\alpha\beta}   \right)  \partial_E f,\\
L_{\alpha\beta\gamma}&=  \frac{e}{h}  \int    dE  \frac{E_F-E}{T}    \left(\frac{\partial A_{\alpha\beta}}{\partial \theta_\gamma }   + \frac{\partial A_{\alpha\gamma}} {\partial \theta_\beta} 
 + \delta_{\beta\gamma} \frac{E-E_F}{T} \frac{\partial A_{\alpha\beta}}{\partial E}  \right) \partial_E f,\\
M_{\alpha\beta\gamma}&= \frac{e^2}{h}  \int  dE \left( \frac{E_F-E}{eT} \frac{\partial A_{\alpha\gamma}}{\partial V_\beta }   -  \frac{\partial A_{\alpha\beta}}{\partial \theta_\gamma}  - \delta_{\beta\gamma}\frac{E-E_F}{T} \frac{\partial A_{\alpha\beta}}{\partial E} \right)\partial_E f \,,\\
R_{\alpha\beta\gamma}&= \frac{e^2}{h} \int dE  \Biggr\{\delta_{\alpha\gamma} A_{\alpha\beta} +\delta_{\alpha\beta} A_{\alpha\beta}
- (E-E_F)\left(\frac{\partial A_{\alpha\beta}}{\partial eV_\gamma}+\frac{\partial A_{\alpha\gamma}}{\partial eV_\beta}\right)-
\delta_{\beta\gamma} \left[(E-E_F)\frac{\partial A_{\alpha\beta}}{\partial E}+A_{\alpha\beta} \right]\Biggr\}\partial_E f,\\ 
K_{\alpha\beta\gamma}&=\frac{1}{h} \int dE \frac{(E-E_F)^2}{T}\Biggr\{\left(\frac{\partial A_{\alpha\beta}}{\partial \theta_\gamma} +\frac{\partial A_{\alpha\gamma}}{\partial \theta_\beta}\right)+ \delta_{\beta\gamma} \left[\frac{(E-E_F)}{T}\frac{\partial A_{\alpha\beta}}{\partial E}+ \frac{A_{\alpha\beta}}{T}\right]\Biggr\}\partial_E f \,, \\ 
 H_{\alpha\beta\gamma}&=\frac{e}{h}\int dE  (E-E_F) \Biggr\{
\left( \frac{\partial A_{\alpha\gamma}}{\partial\theta_\beta}+ \frac{E-E_F}{T}\frac{\partial A_{\alpha\beta}}{\partial eV_\gamma}-\delta_{\alpha\gamma}\frac{A_{\alpha\beta}}{T} \right)+\delta_{\beta\gamma}\left[\frac{(E-E_F)}{T} \frac{\partial A_{\alpha\beta}}{\partial E} + \frac{A_{\alpha\beta}}{T} \right]\Biggr\}\partial_E f\,. 
\end{align}
\end{subequations}
Since the $A$ functions are defined in terms of the scattering matrix,
$s_{\alpha\beta}=s_{\alpha\beta}(\{V_\gamma\},\{\theta_\gamma\},E)$,
which depends on the set of applied voltage and temperature shifts, we need a theory that determines
the variations of the screening potential to changes in the electrochemical potentials and temperatures. To this end,
we expand $U$ up to first order in the driving fields,
\begin{equation}\label{eq_u}
\mathcal{U}=\mathcal{U}_\text{eq} + \sum_\alpha u_\alpha V_\alpha +\sum_\alpha z_\alpha \theta_\alpha\,.
\end{equation}
$u_\alpha=(\partial \mathcal{U}/\partial V_\alpha)_\text{eq}$ is a potential susceptibility (characteristic potential)
that measures the fluctuations of the screening potential to voltage biases~\cite{but93,chr96}. On the other hand,
$z_\alpha=(\partial \mathcal{U}/\partial \theta_\alpha)_\text{eq}$ is the corresponding potential susceptibility
for temperature differences~\cite{san13}. In a mean-field approximation, both potentials satisfy the differential equations
\begin{align}
-\nabla^2 u_\alpha + 4\pi e^2 \Pi u_\alpha &= 4\pi e^2 D^{p}_\alpha \,,\\
-\nabla^2 z_\alpha + 4\pi e^2 \Pi z_\alpha &= 4\pi e D^{e}_\alpha\,.
\end{align}
Here, $\Pi$ is the Lindhard polarization function and $D^{p}_\alpha$ is the particle injectivity,
\begin{equation}\label{eq_partinj}
\nu^{p}_\alpha(E)=\frac{1}{2\pi i}\sum_{\beta}
\Tr\left[ s^\dagger_{\beta\alpha}\frac{d s_{\beta\alpha}}{dE} \right]\,.
\end{equation}
The particle injectivity represents the density of states associated to those carriers that are incoming
from lead $\alpha$ independently of the receiving contacts. Finally, $D^{e}_\alpha$ is the entropic injectivity,
\begin{equation}\label{eq_entrinj}
\nu^{e}_\alpha(E)=\frac{1}{2\pi i}\sum_{\beta}\Tr\left[\frac{E-E_F}{T}
 s^\dagger_{\beta\alpha}\frac{d s_{\beta\alpha}}{dE} \right]\,.
\end{equation}
Since $(E-E_F)/T$ is the entropy of a carrier with energy $E$ relative to the Fermi energy,
this contribution describes a weighted density of states that enters both the heat transfer and the thermoelectric
current beyond linear response. The sum of Eqs.~\eqref{eq_partinj}
and~\eqref{eq_entrinj} allows us to calculate the bare charge
injected into the sample from lead $\alpha$ either electrically or thermally.
The theoretical description is closed by making in Eq.~\eqref{eq_nonlinear} the replacements
\begin{align}\label{replacement}
\partial_{\theta_\gamma} A_{\alpha\beta}=z_\gamma\delta A_{\alpha\beta}/\delta \mathcal{U}&\to -e z_\gamma\partial_E A_{\alpha\beta}\,,\\ \nonumber
\partial_{V_\gamma} A_{\alpha\beta}=u_\gamma\delta A_{\alpha\beta}/\delta \mathcal{U}&\to -e u_\gamma\partial_E A_{\alpha\beta}\,,
\end{align}
which are valid in the WKB approximation (long wavelength limit).

This nonlinear theory of thermoelectric transport can be applied to a variety of situations.
We~\cite{san13} compute the current--temperature characteristics of a quantum dot and find nonlinear contributions
to the thermopower. Interestingly, these contributions are nonvanishing in the zero temperature limit unlike the linear
coefficient. Heat rectification effects and violations of the Wiedemann-Franz law in resonant tunnel-barrier systems
were reported in Ref.~\cite{lop13}.
Meair and Jacquod~\cite{mea13} find that the thermoelectric efficiency in the nonlinear regime differs from linear-response
theory predictions for heat engines and refrigerators based on quantum dot setups and quantum point contacts.
Reference~\cite{aze14} analyzes the range of validity of a linear theory and the conditions under which
the thermal nonlinear terms should be taken into account.
Hwang \textit{et al}.~\cite{hwa13} discuss departures of the Onsager relations in the nonlinear responses that were observed
in the experiment~\cite{mat14}. These magnetic-field asymmetries are already present in the isothermal case,
as shown theoretically~\cite{san05,spi04} and experimentally~\cite{mar06,let06,zum06,ang07,har08}.
The reason is that the nonequilibrium potential need not be a symmetric
function of the magnetic field and the asymmetry is precisely proportional to the electron-electron interaction strength.
The Onsager symmetries can be recovered in chiral quantum Hall antidot systems but the effect is not universal
and depends on the gate potential.
The symmetry breakdown has been also discussed in nanowires in connection with the efficiency of thermoelectric conversion~\cite{cim14}.

\section{Quantum dots}\label{sec-QD}
Quantum dots are electron droplets that confine the charges in three dimensions, thereby enlarging the strength
of electronic correlations. It is thus expected that nonlinear effects and interaction induced phenomena are more visible
in these systems. One of the first thermoelectric experiments indeed reported an interesting observation.
When the temperature difference applied across a two-terminal quantum dot increased,
the created thermovoltage $V_{\rm th}$ also increased as expected from the linear Seebeck effect. Nonetheless,
for more intense heating currents (implying larger temperature bias) $V_{\rm th}$ was observed to decrease
down to zero, even becoming negative. The phenomenon was much more clearly analyzed by Fahlvik Svensson
\textit{et al}.~\cite{sve13,svi15}.
This observation is remarkable and has no analog in the purely electric case since for finite voltage the current is always
nonzero. Here, for a large thermal gradient the thermoelectric current simply stops to flow due to an exact cancellation
of charges flowing in opposite directions far from equilibrium. Small deviations from linearity as discussed in the previous
section cannot account alone for the strongly nonlinear behavior. Therefore, a fully nonlinear model is called for.

A single-level quantum dot with intradot Coulomb interaction can be described with the aid of the Anderson
Hamiltonian,
\begin{equation}\label{eq_ham}
\mathcal{H}=
\sum_{\alpha k \sigma} \varepsilon_{\alpha k} c^\dagger_{\alpha k \sigma} c_{\alpha k \sigma}
+\sum_\sigma \varepsilon_d d_\sigma^\dagger d_\sigma+U d_\uparrow^\dagger d_\uparrow
+\sum_{\alpha k \sigma}(V_{\alpha k}c_{\alpha k \sigma}^\dagger d_\sigma + \rm{h. c.})\,,
\end{equation}
where $c^\dagger_{\alpha k \sigma}$ ($c_{\alpha k \sigma}$) is the creation (annihilation) 
operator for a conduction band electron in lead $\alpha=L,R$ with wavevector $k$ and spin
index $\sigma=\{\uparrow,\downarrow\}$ and $d^\dagger_{\sigma}$ ($d_{\sigma}$) creates (annihilates)
an electron with spin $\sigma$ in the quantum dot. In Eq.~\eqref{eq_ham}
$\varepsilon_{\alpha k}$ is the energy dispersion in the fermionic reservoirs,
$\varepsilon_d$ is the position of the dot energy level (which can be tuned with a nearby gate),
$U$ is the dot charging energy and $V_{\alpha k}$ is the tunnel coupling that hybridizes
dot and leads' states. Within standard many-body techniques and in the wide band limit, the electric and heat currents
are derived~\cite{sie14}
\begin{align}
I&=-\frac{e}{\pi\hbar} \sum_\sigma \frac{\Gamma_L\Gamma_R}{\Gamma} \int dE\, \text{Im}\,
G^r_{\sigma , \sigma}(E)[f_L(E)-f_R(E)]\,,\label{eq_Idot}\\
J_L&=-\frac{1}{\pi\hbar}\sum_\sigma \frac{\Gamma_L\Gamma_R}{\Gamma}\!\!\int\!\! dE \,(E-\mu_L)\text{Im}\,
G^r_{\sigma , \sigma}[f_L(E)-f_R(E)]\,,
\end{align}
in terms of the Fourier transform of the dot retarded Green's function,
$G^r_{\sigma , \sigma}(t)=-i\theta(t)\langle[d_{\sigma}(t),d_\sigma^\dagger(0)]_+\rangle$,
and the tunnel broadenings $\Gamma_\alpha=2\pi \sum_k \delta(E-\varepsilon_{\alpha k})|V_{\alpha}|^2$.
Similarly to the scattering approach where the transmission depends on the applied voltage and temperature
shifts, here the Green function itself is a function of energy and the driving fields,
$G^r=G^r(E,\{V_\alpha\},\{\theta_\alpha\}))$, despite the fact that the interaction energy in the Anderson
model is constant. The reason is that the Green function, even in the simplest decoupling procedure,
depends on the mean occupation in the dot, which is a function of the electrochemical potentials
in the leads and their temperatures. The problem then has to be solved self-consistently.

\begin{figure}[t]
\center
\includegraphics[angle=-90,width=0.9\textwidth,clip]{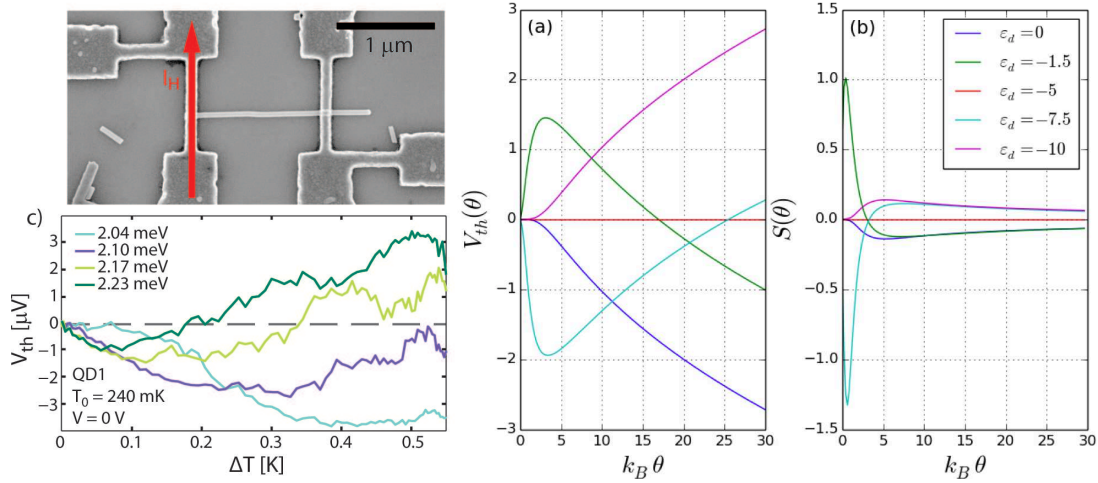}
\caption{(Color online.) Top left: Picture of a nanowire quantum dot device coupled to two external
reservoirs. A heating current $I_H$ traverses the left terminal while the right contact remains cold.
(a) Theoretical calculation of the thermovoltage generated in response to a large temperature difference $\theta$
for various dot level positions. Parameters: charging energy $U=10$ and tunnel broadening $\Gamma=1$.
(b) Nonlinear thermopower of a quantum dot. (c) Experimental thermovoltage measured at different
values of the thermal bias $\Delta T$. Adapted from Refs.~\cite{sve13,sie14}.
}
\label{figdot}
\end{figure}

The experimental setup is depicted in the top left of Fig.~\ref{figdot}~\cite{sve13}. A nanowire quantum dot is attached
to two electrodes with different temperature due to the passage of a heating flow. We then set $I=0$
in Eq.~\eqref{eq_Idot} and find the corresponding thermovoltage $V_{\rm th}$ for a given value
of the thermal bias $\theta$ applied to the left lead. The results are shown in Fig.~\ref{figdot}(a)~\cite{sie14}.
The differential thermopower $S=dV_{\rm th}/d\theta$ is plotted in Fig.~\ref{figdot}(b)
as a function of the dot energy level. We observe that for small $\theta$ the thermovoltage is a linear function
alternating its sign depending on whether electrons flow above or below the Fermi energy.
For increasing $\theta$, both $V_{\rm th}$ and $S$ become strongly nonlinear functions.
For some values of the gate voltage, the thermopower vanishes if temperature is high enough
and reverses its sign. (The particle-hole symmetry point corresponding to $\varepsilon_d=-U/2$ trivially
gives zero current independently of the value of $\theta$). The experimental data [Fig.~\ref{figdot}(c)] show a similar behavior,
which can be ascribed to the presence of two resonances in the
Coulomb-blockaded quantum dot (at energies $\varepsilon_d$
and $\varepsilon_d+U$). Thermally excited electrons from the left reservoir
tunnel through these two channels but one should also take into account the presence of
a flow of holes traveling from right to left that tend to fill the states of the left reservoir below the Fermi energy.
As a consequence, the absolute value of the thermoelectric current
can decrease, attain a nontrivial zero and finally reverse its direction.
The thermovoltage mimics the behavior of the current as shown in both the theory and the experiment.
We emphasize that this is essentially a nonlinear phenomenon without counterpart in the linear regime of transport.
Analogous thermal rectification properties have been reported in the quantum dot literature~\cite{sch08,kuo10,wie10,sta14,sie16}.

\section{Kondo effect}
For temperatures lower than the Kondo temperature $T_K$, the unpaired electron in a quantum dot
forms a many-body spin singlet with the reservoir electrons via higher-order tunneling effects.
The thermopower $S$ serves as a useful spectroscopic tool to probe the correlations that lead to the Kondo effect.
This is so because the linear $S$ is sensitive to the slope of the local density of states.
For instance, the $S$ shows a local minimum for temperatures approaching $T_K$
and changes its sign~\cite{boe01}. In the nonlinear regime, the differential thermopower
defined earlier also changes the sign, indicating a destruction of the Kondo peak
at large temperature differences. Even though the Wiedemann-Franz law is generally not satisfied,
the ratio between the thermal and the electric conductance becomes a constant at very low temperatures
suggesting the complete screening of the dot spin and the emergence of a Fermi liquid picture~\cite{don02}.
Reference~\cite{kra07} shows that the nonlinear thermopower is strongly affected by asymmetric
coupling to the leads unlike the linear $S$ and that in the nonlinear regime the Seebeck coefficient
does not obey any simple scaling law with the asymmetry parameter.
The degenerate double-orbital case is treated in Ref.~\cite{aze12}.
Using a noncrossing approximation out of equilibrium, it is found that power output presents a maximum 
when the left electrode temperature is a few times larger than the right electrode temperature.
This suggests that practical applications should look into the nonlinear regime for optimal efficiencies.
Interestingly, the zero-bias anomaly (Abrikosov-Suhl resonance) connected to the buildup of the
Kondo state breaks particle-hole symmetry due to the interplay
of large temperature gradients and potential biases~\cite{dut13}.
Finally, while in the Coulomb-blockade regime the thermoelectric current can reach a nontrivial zero
as discussed in Sec.~\ref{sec-QD}, the Kondo state leads to multiple nontrivial zeros
and sign reversals due to the splitting of the Kondo peak at finite voltage bias~\cite{zim15}.

\section{Molecular junctions}

\begin{figure}[t]
\center
\includegraphics[angle=-0,width=0.9\textwidth,clip]{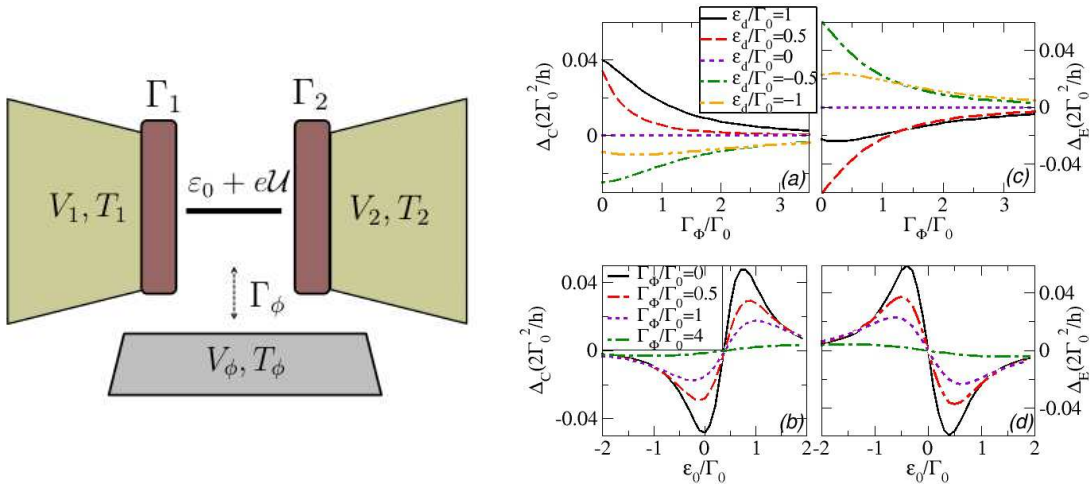}
\caption{(Color online.) Left panel: Sketch of a single-level molecule attached to two reservoirs
of different voltages and temperatures. The nonequilibrium electrostatic potential in the molecule is
denoted with $\mathcal{U}$. Incoherent scattering is modeled with a fictitious probe
whose voltage $V_\phi$ and temperature $T_\phi$ are determined by imposing that the net charge and heat
 currents flowing through the probe
are zero. Then, an electron which enters the probe is reinjected into the molecule with unrelated
phase and randomized energy.
Right panel: (a) Heat current contact asymmetry in the nonlinear regime for increasing coupling
to the probe and (b) as a function of the energy level position tuned with an external gate.
(c) Heat current electric asymmetry
versus the probe coupling and (d) the dot level. Parameters: tunnel broadenings $\Gamma_1=0.1\Gamma_0$ and $\Gamma_2=0.9\Gamma_0$, bias voltage $eV=0.01\Gamma_0$, temperature $k_B\theta=0.01\Gamma_0$,
in the strongly interacting case (capacitance $C=0$).
Adapted from Ref.~\cite{arg15}.
}
\label{figheat}
\end{figure}

When the investigated nanostructure is an atomic chain or a molecular complex, one can reach
an ultimate degree of miniaturization that allows for fundamental studies of charge transport
and heat dissipation. Lee \textit{et al}.~\cite{lee13} built a thermocouple probe attached to single molecules
and Au atomic junctions and measured the heat at the probe and the substrate in the large voltage
bias regime. Quite generally, we can define a contact asymmetry $\Delta_C$, which quantifies
the difference between the heat fluxes in the two terminals, and an electric asymmetry,
which measures the heat current asymmetry at a fixed reservoir swapping the voltage biases
$V_1$ and $V_2$:
\begin{align}
\Delta_C &=J_1(V_1,V_2)-J_2(V_1,V_2)\,,\\
\Delta_E &=J_1(V_1,V_2)-J_1(V_2,V_1)\,.
\end{align}
To lowest order in a voltage expansion, both asymmetries agree ($\Delta_C=\Delta_E\equiv \Delta$)
and $\Delta$ is proportional to the thermopower of the junction.
Therefore, molecules with asymmetric transmission lineshapes experimentally
exhibit this phenomenon while
the dissipated power is independent of the bias polarity in Au chains with transmissions showing a weak 
energy dependence. Reference~\cite{zot14} goes beyond the linear regime and finds that Joule
heating dominates over Peltier cooling for most of the voltage range.
The electrostatic control of thermoelectricity using a gate voltage is achieved in Ref.~\cite{kim14},
where thermal gradients as extreme as $10^9$~K/m are reported.
Reference~\cite{arg15} discusses dephasing and inelastic effects in the linear and nonlinear
heat through a molecular transistor. The subject is important because molecules usually display
vibrational modes that not only carry energy but also interact with the conduction electrons~\cite{koc14}. A phenomenological
model using fictitious voltage probes is proposed (see left panel in Fig.~\ref{figheat}).
Notably, the heat-current asymmetry
is still found to be given by a renormalized thermopower that takes into account the 
transitions into and out of the probe (inelastic scattering). This relation holds at very low temperatures
in which case a leading-order Sommerfeld expansion is valid. In the nonlinear case, the heat flux
must include electron-electron interaction as discussed in Sec.~\ref{sec-SA}.
When the molecule is described with a double-barrier resonant model and a fluctuating energy level,
rectification effects, $J_L(V)\neq J_L(-V)$, are observed in the heat--voltage characteristics.
More interestingly, particle-hole symmetry is broken 
in the nonlinear regime and for a small amount of incoherent scattering, see Fig.~\ref{figheat}(a).
This breaking does not occur in the electrical asymmetry [Fig.~\ref{figheat}(c)],
which demonstrates that both asymmetries
differ beyond linear response. Both heat asymmetries, however, show a double peak
structure as a function of the gate voltage [see Figs.~\ref{figheat}(b) and~(d)].

Purely thermal rectification have been also predicted in molecular junctions~\cite{seg05,seg06}.
Based on a master equation approach, negative differential thermal
conductance is found depending on the coupling to the thermal baths, which are described
with a set of independent harmonic oscillators, and the presence of strongly
anharmonic vibrational modes in the molecule. Leijnse \textit{et al}.~\cite{lei10} also consider
the charge currents within a similar theoretical framework and calculate the efficiency
and the output power in the nonlinear regime. Maximal efficiency is reached when
the level lies far above the Fermi energy in the leads. A combination of the scattering matrix formalism
with the nonequilibrium Green's function approach also provides encouraging results
for the nonlinear conversion of heat into useful work employing molecular bridges
with electron-electron and electron-phonon interactions~\cite{zim15b}.

\section{Heat rectification and nonlinear Peltier effects}
Heat rectifying mechanisms do not take place only in molecular junctions.
For instance,  Chang \textit{et al}.~\cite{cha06} demonstrate thermal rectification
in carbon nanotubes. However, in this work it is claimed that the responsible mechanism
are solitonic waves of phonons.
This is an interesting path of research but falls outside the scope of the present
paper since we are here mainly interested in the electronic contribution to heat flows.
An example is Ref.~\cite{ruo11}, which proposes a heat diode based on Coulomb
interactions between serially coupled double quantum dots. As a result of the combined effect of
Coulomb blockade and reservoirs with different temperatures, heat flow
strongly depends on the transport direction.
Jiang \textit{et al}.~\cite{jia15} consider the coupling between electronic and phononic degrees
of freedom in the heat currents of a double quantum dot system.
They find that inelastic processes lead
to rectification effects and thermal transistor behaviors.
The case of a purely electronic quantum dot is treated in Ref.~\cite{sie15}.
It is argued that in the voltage-driven case the Joule term quickly surpasses the Peltier heat.
Departures from Fourier's law are found in the thermal current, which monotonically
increases with temperature difference $\theta$. In fact, the differential thermal conductance
scales as $1/\theta$ for large driving fields $\theta$, showing a great sensitivity to gate voltages.
In general, Fourier's law is generally violated in nanoscale junctions~\cite{dub09}
Furthermore, the Kelvin-Onsager relation is shown to break down in the nonlinear
regime, the deviation being larger for positions of the dot level far away from the
electron-hole symmetry point. 
These results are valid when sequential tunneling processes are dominant.
However, when the background temperature is of the order or lesser than
the coupling to the leads, cotunneling mechanisms take over and charge fluctuations should
then be taken into account. Gergs \textit{et al}.~\cite{ger15} consider nonlinearities in
the heat transport due to virtual occupation of the charge states and predict
a strong negative differential heat conductance that provides spectroscopic
information about relaxation processes mostly invisible to the electric current.
Therefore, general discussions on theoretical formalisms for
nonlinear quantum heat~\cite{yam15,sel16}
are needed to pinpoint the relevance of heat fluxes in applications
and fundamental microscopic research.

The first works that investigated nonlinear cooling effects due to higher-order Peltier
heat formulated semiclassical models suitable for small systems.
In Ref.~\cite{kul94} a kinetic equation is solved for the electronic distribution function $f$
of a metallic microconstriction
beyond linear response. The leading-order nonlinearity in the Peltier coefficient resembles 
an effective temperature that dominates at low $T$.
The theory assumes a constant relaxation time approximation. A 
Zeberjadi \textit{et al}.~\cite{zeb07}
perform a Monte Carlo simulation for the Boltzmann transport equation governing
electron dynamics in doped bulk semiconductors
and find that nonlinear Peltier effect can cool thin film devices at low temperatures.
Thin films are also the subject of interest in Ref.~\cite{gri90},
which obtains a thermopower expansion in powers of the temperature difference
which depends on the spatial direction.
Quantum point contacts are considered in Ref.~\cite{bog99}.
Deviations from the Onsager relations are naturally found in the nonlinear regime. Additionally,
the differential Peltier coefficient presents oscillations induced by an external magnetic field
with increasing applied bias voltage.
The experimental measurement of the nonlinear thermopower of a one-dimensional ballistic
system reveals that, contrary to expectations, a linear-response model explains well the results~\cite{dzu93}.
Returning to the Peltier effect, Whitney~\cite{whi13} predicts 
that, for driving currents above a critical value, nonlinear cooling of point contacts can be achieved
down to zero temperature.

\section{Efficiencies in the nonlinear regime}
As discussed above Eq.~\eqref{eq_I}, the widely used Ioffe formula for the thermoelectric figure of merit
depends on linear responses only~\cite{ioffe}. Therefore, new methods to calculate efficiencies in the
nonlinear realm must be devised. Reference~\cite{whi14} asks about the maximum efficiency
of a quantum heat engine or refrigerator working with finite power output.
The answer is provided by nonlinear scattering theory: there exists an upper bound
to this efficiency, which can be attained in systems with 'boxcar' transmission functions
as in chains of coupled quantum dots or molecules.
The bounds influence nonlinear heating, work and entropy production~\cite{whi13b}
and are connected to the first and the second laws of thermodynamics.
Moreover, the quantum bound on the heat current out of a particular reservoir
limits the performance of the engine. However, the power output can be scaled up by connecting
side-coupled quantum dots without lowering the efficiency~\cite{her13}.
The optimal transmission probability is not just a rapidly varying function around
the Fermi energy as in the linear regime but needs to include the different Fermi functions
in the leads, and is thus dependent on the applied temperature gradient and voltage bias.
An alternative is to use gated semiconducting nanowires, which show high efficiency
in the presence of disorder~\cite{mut15}. The transmission can be then appropriately tuned
with a gate voltage to be large for certain energy ranges. 
Three-terminal setups have also practical advantages due to a spatial separation
of hot and cold baths that facilitates a reduction fo leakage heat currents~\cite{sot15}.
In the nonlinear regime, the maximal output power and efficiency remain without modification
despite nonequilibrium screening can affect the optimal distace between the dot energy levels~\cite{szu16}.

\section{Miscellaneous topics: superconductors, topological insulators, noise}
\begin{figure}[t]
\center
\includegraphics[angle=-0,width=0.7\textwidth,clip]{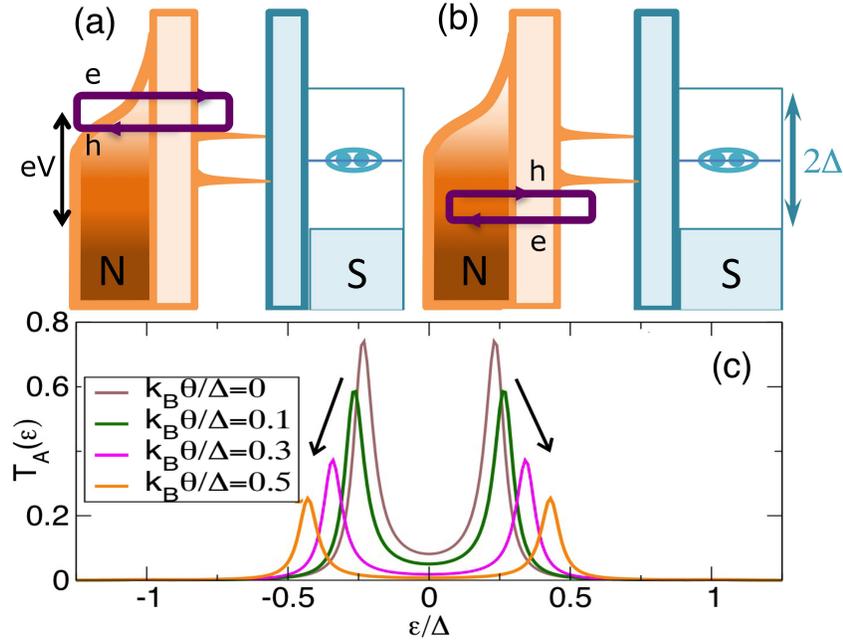}
\caption{(Color online.) (a) and (b) Sketch of Andreev processes taking place
in a quantum dot tunnel coupled to a hot, normal lead ($N$) and a cold,
superconducting terminal ($S$). A thermally smeared Fermi distribution
is depicted for the $N$ side, which is electrically biased with voltage $V$.
$2\Delta$ is the size of the superconductor gap. (c) Andreev transmission
as a function of energy for different values of the thermal gradient $\theta$.
Note the temperature induced shift of the bound states out of equilibrium.
Reproduced from Ref.~\cite{hwa15}.
}
\label{figsuper}
\end{figure}

Thus far, we have considered normal leads and fluxes that do not fluctuate in time.
Very little is known when more exotic materials (superconductors, topological
insulators) or noise correlations come into play beyond linear response.
We now give a brief account of the latest developments.
We begin with the superconducting case. Nonlinear thermoelectric currents
were investigated in Ref.~\cite{hwa15}. A quantum dot sandwiched between
a normal lead and a superconducting electrode offers a unique opportunity
to examine cross effects. Since Andreev processes [the retroreflection
of an electron as a hole at the superconducting interface or vice versa,
see Fig.~\ref{figsuper}(a) and~(b)] depend on the Fermi distribution
of the normal lead only, the Andreev current disappears to all orders
in the temperature bias and no thermopower can then be generated
due to the particle-hole symmetric nature of Andreev processes.
However, in the nonlinear regime of transport the cross term described
by the $M_{\alpha\beta\gamma}$ coefficient in Eq.~\eqref{currents}
is nonzero. As a consequence, the Andreev transmission, which shows
a typical double peak structure as in Fig.~~\ref{figsuper}(c),
becomes shifted as the temperature bias applied to the normal lead
increases (the superconducting side remains cold).
These effects dominate in the subgap regime. For larger energies,
quasiparticle currents are to be considered and the pure thermoelectric
response can now be finite. In fact, the quasiparticle contribution
is largest at finite voltages, which boosts the nonlinear thermovoltage
when the coupling to the superconducting reservoir is greater than
the coupling to the normal metal. Hybrid systems are also relevant
when one examines the heat currents. A thermal diode has been
experimentally realized in a normal-insulator-superconducting junction
with a highly efficient rectification parameter~\cite{mar15}.
The working principle of this device is the strong temperature dependence
of the superconducting density of states~\cite{gia13}.
Strikingly, thermal rectifiers with extremely high rectification values
have been proposed in similar tunnel-junction setups without superconductors~\cite{for14}.
The key point here is to have three metallic islands and the central
electrode coupled to a phonon bath. 

\begin{figure}[t]
\center
\includegraphics[angle=-0,width=0.99\textwidth,clip]{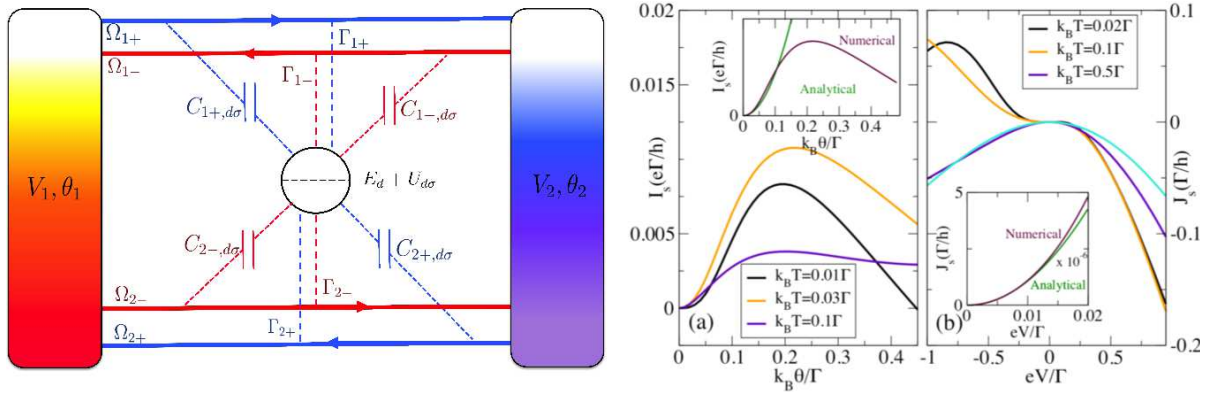}
\caption{(Color online.) Left: Quantum spin Hall conductor attached to
two terminals ($1$ and $2$) with differing voltage and temperature shifts.
An single-level antidot (circle) is located at the middle of the sample.
The antidot is coupled to helical edge states (coupling constant $\Gamma$),
where $s=\pm$ labels the helicity and $\sigma$ the spin. Interactions arise
from capacitive interactions (denoted wih $C$) between the resonant
level and the edges. Right: (a) Pure spin charge current as a function of the applied
thermal gradient ($\theta_1=\theta$, $\theta_2=0$) for different values
of the background temperature $T$. Inset: comparison between analytical
and exact numerical calculations in the low gradient regime. (b) Pure spin heat
current versus voltage bias ($V_1=V$, $V_2=0$). The inset highlights
the quadratic dependence of the heat as a function of a small applied voltage.
Adapted from Ref.~\cite{hwa14}. 
}
\label{figQSH}
\end{figure}

Two-dimensional topological insulators exhibit quantum spin Hall
effect due to the existence of gapless edge states that exhibit spin-momentum
locking: opposite spins belonging to the same channel propagate
in opposite directions~\cite{ber06,kon07}. Quantum spin Hall conductors
with artificial backscattering centers would provide spin-filtering effects.
An illustration is shown in the left panel of Fig.~\ref{figQSH}. The topological insulator
is connected to two terminals ($1$ and $2$) and the helical edge states
are plotted with straight lines with different colors depending
on the spin direction. In the center of the sample, a potential hill
(antidot) can be formed with external gates. It supports a resonant
level that deflects electrons between the upper and lower edge states.
We also consider capacitive couplings between the dot and the edge states.
In linear response and in the limit of vanishingly small couplings to the dot,
the system conductance is quantized as $2e^2/h$.
In the nonlinear regime, electron-electron interactions are treated
within a Hartree (mean-field) approach~\cite{hwa14}.
Remarkably, the nonequilibrium potential becomes spin dependent
when the dot is asymmetrically coupled to the edges since
more electrons with a given spin direction are injected with a finite
voltage bias. Hence, a spin-polarized current for both charge and heat
emerges but only in the nonlinear regime. 
This spin polarization can be generated with thermal means only
and vanishes in the zero temperature limit as the Seebeck effect.
Interestingly, pure spin currents $I_s$ and $J_s$ can be created for zero charge
currents. The latter constraint can be accomplished with a thermovoltage
in response to a temperature bias $\theta$. 
The numerical calculation is shown in Fig.~\ref{figQSH}(a) and~(b).
The inset shows a comparison between the numerical and the analytical results.
In both cases, the leading order goes as $\theta^2$. Therefore, the effect
is purely nonlinear. Figure~\ref{figQSH}(b) shows the pure spin heat current
as a function of the bias voltage. Here, a thermal bias should be generated
to adiabatically isolate the conductor. $J_s$ is also quadratic in $V$ at low voltages
(see inset),
again emphasizing the nonlinear origin of the spin polarization mechanism.
The asymmetric coupling of the antidot
is an essential ingredient of the spin-filter effect~\cite{dol13}.
Furthermore, the sign of the spin-polarized current can be
manipulated with a change of the voltage bias and the phenomenon
is robust against the presence of interactions in the bar edges.
For magnetic contacts, the unequal spin injection from the terminals
already generates spin-polarized currents, which shows as an additional
contribution to the interaction induced effect~\cite{lop14}. This, however, only
takes place when the magnetic moments of the leads are in a parallel
configuration because in the antiparallel case
the linear conductance coefficients become spin independent
due to to symmetry relations. 
A pure spin current can be also generated in a serially coupled double quantum point
contact built on a two-dimensional topological insulator
that is attached to two reservoirs in the presence of a thermal gradient~\cite{ron16}.
In contrast to the charge current,
the spin current arises in the nonlinear regime only and it can be tuned with
a gate voltage or changing the distance between the point contacts.
Interference effects due to scattering from multiple quantum point contacts
in fractional quantum Hall systems are investigated in Ref.~\cite{van15}.
The resulting heat current can rectify if two systems with different filling
factors become coupled. 

Dynamical fluctuations of currents is a fertile area of research
in mesoscopic physics~\cite{bla00}. Fluctuations can be due to thermal agitation
or to the quantization of charge and can reveal additional
information about the internal energy scale and dynamics
of an electronic nanostructure. For purely charge transport,
noise power is expressed as a current-current correlation function.
In the presence of heat flow, one now also has the possibility
to examine cross correlations between the charge and the heat
currents. For a three-terminal quantum-dot energy harvester
it is found that charge-heat fluctuations are maximal
for optimal configurations where Carnot efficiency reaches a maximum value~\cite{sanc13}.
Moreover, transitions between sub- and super-Poissonian noise are identified.
The mutual correlation between charge and heat fluctuations
is explored in Ref.~\cite{cre15} for both the linear and nonlinear regimes.
While the linear response case is characterized by equal ratios
between the mixed noise with respect to the product of charge and heat noises,
in the nonlinear regime these ratios differ, pointing out another fundamental
contrast between the two regimes of transport.
The noise can be generalized to all orders using a cumulant-generating function,
which fulfills a fluctuation theorem. An expansion in the affinities reveals
that the nonlinear conductances are connected to the linear-response noise
to leading order~\cite{for08,sai08,san09,esp09,nak10,lop12}.
For the case of heat transfer, Ref.~\cite{uts14}
shows, after integrating out the energy fluctuations of a probe electrode,
that the two-terminal heat-heat correlator is also related to the nonlinear
thermal conductance.

\section{Summary and outlook}

Although the field of nonlinear quantum thermoelectrics and heat is still in its infancy,
it has already produced a number of remarkable results. Understanding the flow of electrons
in mesoscopic and nanoscale systems beyond linear response requires a detailed knowledge
of the nonequilibrium interaction properties of carriers confined to low dimensions.
In particular, symmetry laws fulfilled in the linear regime do not necessarily hold far from equilibrium.
Rectification effects of a pure thermal origin arise in temperature driven charge and heat currents.
Furthermore, the nonlinear thermoelectric current can vanish in quantum dot systems
and reverse its sign for a large temperature
difference. These sign changes are also observed in the differential thermopower both in the Coulomb
blockade and in the Kondo regimes, suggesting that the nonlinear Seebeck coefficient can be used
as a complementary probe to study electronic correlations in nanostructures.
The study of heat dissipation in molecular junctions have both fundamental and practical motivations.
Asymmetric dissipation even in the presence of inelastic scattering occurs in the isothermal
case whereas negative differential thermal conductances might be possible in molecular bridges.
Heat rectification exclusively due to electron transfer may give rise to the implementation of thermal
diodes for cooling devices and thermal information processing. Thermoelectric applications
beyond linear response need a clear understanding of the efficiencies far from equilibrium.
Steps in this direction have been given. Less is known about the influence of superconductivity
and topological insulators. However, recent studies show the thermoelectric tuning
of Andreev and spin-dependence transmissions using large temperature gradients. Furthermore,
the simultaneous presence of charge and heat current naturally leads to interest in mixed noises
with characteristic properties in the nonlinear regime of transport.

Rather than trying to now give an exhaustive list of open problems, we will just outline a few future prospects
that, in our opinion, might become important. It is clear from our discussion that theory has advanced
despite the inherent difficulty of the nonlinear regime and that there are a number of interesting predictions
that should be tested experimentally.
However, many of the obstacles for a rapid development of the field
stem from the fact that the experimental manipulation of heat flows and accurate measurements
of temperature gradients remain challenging.
Also, from the theoretical side there are scarce works that go beyond the weakly nonlinear regime.
To the best of our knowledge, theoretical discussions have been mostly focused on zero-frequency
dc currents. Dynamics of heat and thermoelectric currents is an exciting subject that should also
be treated beyond the adiabatic (small perturbations) regime. 
The question of strong electronic correlations is also open. Further investigations should consider
the synergistic participation of electron-phonon and electron-photon interaction in the transport beyond
linear response. The role of disorder has not been addressed either. One might also envisage
the generation of self-sustained oscillations and chaos as prototypical nonlinear phenomena
but driven by purely thermal means. This plethora of possibilities makes us foresee a field full of rewards
and nice opportunities for the near future.




\section*{Acknowledgements}
This work was supported by MINECO under Grant No.\ FIS2014-52564.


\begin{thebibliography}{00}
\bibitem{sze}
S. M. Sze, \textit{Physics of Semiconductor Devices} (Wiley, New York, 1981).
\bibitem{esaki}
L. Esaki, Phys. Rev. \textbf{109}, 603 (1958).
\bibitem{gunn}
J. B. Gunn, Solid State Commun. \textbf{1}, 88 (1963).
\bibitem{onsager}
L. Onsager, Phys. Rev. {\bf 38}, 2265 (1931).

\bibitem{ven01}
R. Venkatasubramanian, E. Siivola, T. Colpitts, and B. O'Quinn, Nature \textbf{413}, 597 (2001).
\bibitem{red07}
P. Reddy, S. Y. Jang, R. A. Segalman, and A. Majumdar,
Science \textbf{315}, 1568 (2007).

\bibitem{vin10}
C. J. Vineis, A. Shakouri, A. Majumdar, and M. G. Kanatzidis,
Adv. Mater. \textbf{22}, 3970 (2010).
\bibitem{dub11}
Y. Dubi and M. Di Ventra,
Rev. Mod. Phys. \textbf{83}, 131 (2011).
\bibitem{here13}
J. P. Heremans, M. S. Dresselhaus, L. E. Bell, and D. T. Morelli,
Nature Nanotechnology \textbf{8}, 471 (2013).
\bibitem{san13b}
D. S\'anchez and H. Linke, New J. Phys. \textbf{16}, 110201 (2014).
\bibitem{zim16}
N. A. Zimbovskaya, J. Phys.: Condens. Matter \textbf{28}, 183002 (2016).
\bibitem{book}
\textit{Thermoelectrics Handbook. Macro to Nano}, edited by D. M. Rowe (CRC Press, Boca Raton, FL, 2006).

\bibitem{ter02}
M. Terraneo, M. Peyrard, and G. Casati, Phys. Rev. Lett.
\textbf{88}, 094302 (2002).

\bibitem{san13}
D. S\'anchez and R. L\'opez, 
Phys. Rev. Lett. \textbf{110}, 026804 (2013).
\bibitem{lop13}
R. L\'opez and D. S\'anchez, 
Phys. Rev. B \textbf{88}, 045129 (2013).

\bibitem{siv86}
U. Sivan and Y. Imry, Phys. Rev. B \textbf{33}, 551 (1986).
\bibitem{but90}
P. N. Butcher, J. Phys.: Condens. Matter \textbf{2}, 4869 (1990).

%
\bibitem{but86}
M. B\"uttiker, Phys. Rev. Lett. \textbf{57}, 1761 (1986).
\bibitem{ben86}
A. D. Benoit, S. Washburn, C. P. Umbach, R. B. Laibowitz, and R. A. Webb, Phys. Rev. Lett. \textbf{57}, 1765 (1986).
\bibitem{san11}
D. S\'anchez and L. Serra, Phys. Rev. B \textbf{84}, 201307(R) (2011).
\bibitem{sai11}
K. Saito, G. Benenti, G. Casati, and T. Prosen, Phys. Rev. B \textbf{84}, 201306(R) (2011).
\bibitem{bra13}
K. Brandner, K. Saito, and U. Seifert, Phys. Rev. Lett. \textbf{110}, 070603 (2013).
\bibitem{mat14}
J. Matthews, F. Battista, D. S\'anchez, P. Samuelsson, and H. Linke, 
Phys. Rev. B \textbf{90}, 165428 (2014).
\bibitem{mat12}
J. Matthews, D. S\'anchez, M. Larsson, and H. Linke,
Phys. Rev. B \textbf{85}, 205309 (2012).

\bibitem{but93}
 M. B\"uttiker, J. Phys.: Condens. Matter \textbf{5}, 9361 (1993).
\bibitem{chr96}
T. Christen and M. B\"uttiker, 
Europhys. Lett. \textbf{35}, 523 (1996).

\bibitem{mea13}
J. Meair and P. Jacquod, 
J. Phys.: Condens. Matter \textbf{25}, 082201 (2013).
\bibitem{aze14}
J. Azema, P. Lombardo, and A.-M. Dar\'e,
Phys. Rev. B \textbf{90}, 205437 (2014).

\bibitem{hwa13}
S.-Y. Hwang, D. S\'anchez, M. Lee, and R. L\'opez, 
New. J. Phys. \textbf{15}, 105012 (2013).

\bibitem{san05}
D. S\'anchez and M. B\"uttiker, Phys. Rev. Lett. \textbf{93}, 106802
(2004); Int. J. Quantum Chem. \textbf{105}, 906 (2005).
\bibitem{spi04}
B. Spivak and A. Zyuzin, Phys. Rev. Lett. \textbf{93}, 226801 (2004).
\bibitem{mar06}
C. A. Marlow, R. P. Taylor, M. Fairbanks, I. Shorubalko,
and H. Linke, Phys. Rev. Lett. \textbf{96}, 116801 (2006).
\bibitem{let06}
R. Leturcq, D. S\'anchez, G. G\"otz, T. Ihn, K. Ensslin, D. C. Driscoll, and A. C. Gossard,
Phys. Rev. Lett. \textbf{96}, 126801 (2006).
\bibitem{zum06}
D. M. Zumb\"uhl, C. M. Marcus, M. P. Hanson, and A. C.
Gossard, Phys. Rev. Lett. \textbf{96}, 206802 (2006).
\bibitem{ang07}
L. Angers, E. Zakka-Bajjani, R. Deblock, S. Gu\'eron, H. Bouchiat, A. Cavanna, U. Gennser, and M. Polianksi,
Phys. Rev. B \textbf{75}, 115309 (2007).
\bibitem{har08}
D. Hartmann, L. Worschech, and A. Forchel, Phys. Rev. B
\textbf{78}, 113306 (2008).

\bibitem{cim14}
V. A. Cimmelli, A. Sellitto, and D. Jou,
Proc. R. Soc. A \textbf{470}, 20140265 (2014).

\bibitem{sta93}
A. A. M. Staring, L. W. Molenkamp, B. W. Alphenaar, H. van Houten, O. J. A. Buyk, M. A. A. Mabesoone, C. W. J. Beenakker, and C. T. Foxon, Europhys. Lett. \textbf{22}, 57 (1993).
\bibitem{sve13}
S. F. Svensson, E. A. Hoffmann, N. Nakpathomkun, P. M. Wu, H. Q. Xu, H. A. Nilsson, D. S\'anchez, V. Kashcheyevs, and H. Linke, 
New. J. Phys. \textbf{15}, 105011 (2013).
\bibitem{svi15}
A. Svilans, A. M. Burke, S. Fahlvik Svensson, M. Leijnse, and H. Linke,
Physica E \textbf{82}, 34 (2016).
\bibitem{sie14}
M. A. Sierra and D. S\'anchez, 
Phys. Rev. B \textbf{90}, 115313 (2014).
\bibitem{sch08}
R. Scheibner, M. K\"onig, D. Reuter, A. D. Wieck, C. Gould, H. Buhmann, and L. W. Molenkamp,
New J. Phys. \textbf{10}, 083016 (2008).
\bibitem{kuo10}
D. M.-T. Kuo and Y.-C. Chang, 
Phys. Rev. B \textbf{81}, 205321 (2010).
\bibitem{wie10}
M. Wierzbicki and R. \'Swirkowicz,
Phys. Rev. B \textbf{82}, 165334 (2010).
\bibitem{sta14}
A. E. Stanciu, G. A. Nemnes, and A. Manolescu,
Rom. J. Phys. 60, 716 (2015).
\bibitem{sie16}
M. A. Sierra, M. Saiz-Bret\'{\i}n, F. Dom\'{\i}nguez-Adame, and D. S\'anchez,
Phys. Rev. B  \textbf{93}, 224509 (2016).

\bibitem{boe01}
D. Boese and R. Fazio, Europhys. Lett. \textbf{56}, 576 (2001).
\bibitem{don02}
B. Dong and X. L. Lei, J. Phys.: Condens. Matter \textbf{14}, 11747 (2002).
\bibitem{kra07}
M. Krawiec and K. I. Wysoki\'nski, Phys. Rev. B \textbf{75}, 155330 (2007).
\bibitem{aze12}
J. Azema, A.-M. Dar\'e, S. Sch\"afer, and P. Lombardo,
Phys. Rev. B \textbf{86}, 075303 (2012).
\bibitem{dut13}
P. Dutt and K. Le Hur, Phys. Rev. B \textbf{88}, 235133 (2013).
\bibitem{zim15}
N. A. Zimbovskaya, J. Chem. Phys. \textbf{142}, 244310 (2015).

\bibitem{lee13}
W. Lee, K. Kim, W. Jeong, L. A. Zotti, F. Pauly, J. C. Cuevas, and P. Reddy, Nature (London) \textbf{498}, 209 (2013)
\bibitem{zot14}
L. A. Zotti, M. Bürkle, F. Pauly, W. Lee, K. Kim, W. Jeong, Y. Asai, P. Reddy, and J. C. Cuevas, New J. Phys. \textbf{16}, 015004 (2014).
\bibitem{kim14}
Y. Kim, W. Jeong, K. Kim, W. Lee, P. Reddy, Nature Nanotech. \textbf{9}, 881 (2014).
\bibitem{arg15}
J. Arg\"uello-Luengo, D. S\'anchez and R. L\'opez, Phys. Rev. B \textbf{91}, 165431 (2015).
\bibitem{koc14}
T. Koch, J. Loos, and H. Fehske, Phys. Rev. B \textbf{89}, 155133 (2014).
\bibitem{seg05}
D. Segal and A. Nitzan, J. Chem. Phys. \textbf{122}, 194704 (2005).
\bibitem{seg06}
D. Segal, 
Phys. Rev. B \textbf{73}, 205415 (2006).
\bibitem{lei10}
M. Leijnse, M. R. Wegewijs, and K. Flensberg,
Phys. Rev. B \textbf{82}, 045412 (2010).
\bibitem{zim15b}
N. A. Zimbovskaya, Physica E \textbf{74}, 213 (2015).

\bibitem{cha06}
C. W. Chang, D. Okawa, A. Majumdar, and A. Zettl, Science \textbf{314}, 1121 (2006).
\bibitem{ruo11}
T. Ruokola and T. Ojanen, Phys. Rev. B \textbf{83}, 241404 (2011).
\bibitem{jia15}
J.-H. Jiang, M. Kulkarni, D. Segal, and Y. Imry,
Phys. Rev. B \textbf{92}, 045309 (2015).
\bibitem{sie15}
M. A. Sierra and D. S\'anchez,
Materials Today: Proceedings \textbf{2}, 483 (2015).
\bibitem{dub09}
Y. Dubi and M. Di Ventra, Nano Lett. \textbf{9}, 97 (2009).
\bibitem{ger15}
N. M. Gergs, C. B. M. H\"orig, M. R. Wegewijs, and D. Schuricht,
Phys. Rev. B \textbf{91}, 201107(R) (2015).
\bibitem{yam15}
K. Yamamoto and N. Hatano,
Phys. Rev. E \textbf{92}, 042165 (2015).
\bibitem{sel16}
A. Sellitto, V. A. Cimmelli, and D. Jou,
\textit{Mesoscopic Theories of Heat Transport in Nanosystems}
(Springer, Cham, 2016).

\bibitem{kul94}
I. O. Kulik, 
J. Phys.: Condens. Matter \textbf{6}, 9737 (1994).
\bibitem{gri90}
A. N. Grigorenko, P. I. Nikitin, Daniel A. Jelski, and Thomas F. George,
Phys. Rev. B \textbf{42}, 7405 (1990).
\bibitem{zeb07}
M. Zeberjadi, K. Esfarjani, and A. Shakouri,
Appl. Phys. Lett. \textbf{91}, 122104 (2007).
\bibitem{bog99}
E. N. Bogachek, A. G. Scherbakov, and U. Landman, 
Phys. Rev. B \textbf{60}, 11678 (1999).
\bibitem{dzu93}
A. S. Dzurak, C. G. Smith, L. Mart\'{\i}n-Moreno, M. Pepper, D. A. Ritchie, G. A. C. Jones, and D. G. Hasko,
J. Phys.: Condens. Matter \textbf{5}, 8055 (1993).
\bibitem{whi13}
R. S. Whitney, Phys. Rev. B \textbf{88}, 064302 (2013).

\bibitem{ioffe}
A. F. Ioffe, \textit{Semiconductor Thermoelements and Thermoelectric Cooling}
(Infosearch Ltd., London, 1956).

\bibitem{whi14}
R. S. Whitney, Phys. Rev. Lett. \textbf{112}, 130601 (2014); Phys. Rev. B \textbf{91}, 115425 (2015).
\bibitem{whi13b}
R. S. Whitney, Phys. Rev. B \textbf{87}, 115404 (2013).
\bibitem{her13}
S. Hershfield, K. A. Muttalib, and B. J. Nartowt, Phys. Rev. B \textbf{88}, 085426 (2013).
\bibitem{mut15}
K. A. Muttalib, and S. Hershfield, Phys. Rev. Appl. \textbf{3}, 054003 (2015).
\bibitem{sot15}
B. Sothmann, R. S\'anchez, and A. N. Jordan,
Nanotechnlogy \textbf{26}, 032001 (2015).
\bibitem{szu16}
B. Szukiewicz, U. Eckern, and K. I. Wysokin\'nski,
New J. Phys. \textbf{18}, 023050 (2016).

\bibitem{hwa15}
S.-Y. Hwang, R. L\'opez, and D. S\'anchez,
Phys. Rev. B \textbf{91}, 104518 (2015).
\bibitem{mar15}
M. J. Mart\'{\i}nez-P\'erez, A. Fornieri, and F. Giazotto,
Nature Nanotechnlogy \textbf{10}, 303 (2015).
\bibitem{gia13}
F. Giazotto and F. S. Bergeret,
Appl. Phys. Lett. \textbf{103}, 242602 (2013).
\bibitem{for14}
A. Fornieri, M. J. Mart\'{\i}nez-P\'erez, and F. Giazotto,
Appl. Phys. Lett. \textbf{104}, 183108 (2014).

\bibitem{ber06}
B. A. Bernevig, T. L. Hughes, and S.-C. Zhang, Science \textbf{314},
1757 (2006).
\bibitem{kon07}
 M. K\"onig, S. Weidmann, C. Brune, A. Roth, H. Buhmann,
L. W. Molenkamp, X.-L. Qi, and S.-C. Zhang, Science \textbf{318},
766 (2007).

\bibitem{hwa14}
S.-Y. Hwang, R. L\'opez, M. Lee, and D. S\'anchez,
Phys. Rev. B \textbf{90}, 115301 (2014).
\bibitem{dol13}
G. Dolcetto, F. Cavaliere, D. Ferraro, and M. Sassetti,
Phys. Rev. B \textbf{87}, 085425 (2013).
\bibitem{lop14}
R. L\'opez, S.-Y. Hwang, and D. S\'anchez,
J. Phys.: Conf. Ser. \textbf{568}, 052016 (2014).
\bibitem{ron16}
F. Ronetti, L. Vanucci, G. Dolcetto, M. Carrega, and M. Sassetti,
Phys. Rev. B \textbf{93}, 165414 (2016).
\bibitem{van15}
L. Vanucci, F. Ronetti, G. Dolcetto, M. Carrega, and M. Sassetti,
Phys. Rev. B \textbf{92}, 075446 (2015).


\bibitem{bla00}
Ya. M. Blanter and M. B\"uttiker, Phys. Rep. \textbf{336}, 1 (2000).

\bibitem{sanc13}
R. S\'anchez, B. Sothmann, A. N. Jordan, and M. B\"uttiker,
New J. Phys. \textbf{15}, 125001 (2013).
\bibitem{cre15}
A. Cr\'epieux and F. Michelini,
J. Phys.: Condens. Matt. \textbf{27}, 015302 (2015).
\bibitem{for08}
H. F\"orster and M. B\"uttiker, Phys. Rev. Lett. \textbf{101}, 136805 (2008).
\bibitem{sai08}
K. Saito and Y. Utsumi, Phys. Rev. B \textbf{78}, 115429 (2008).
\bibitem{san09}
D. S\'anchez, Phys. Rev. B \textbf{79}, 045305 (2009).
\bibitem{esp09}
M. Esposito, U. Harbola, and S. Mukamel, Rev. Mod. Phys. \textbf{81}, 1665 (2009).
\bibitem{nak10}
S. Nakamura, Y. Yamauchi, M. Hashisaka, K. Chida, K. Kobayashi, T. Ono, R. Leturcq,
K. Ensslin, K. Saito, Y. Utsumi, and A. C. Gossard, Phys. Rev. Lett. \textbf{104}, 080602 (2010);
Phys. Rev. B \textbf{83}, 155431 (2011).
\bibitem{lop12}
R. L\'opez, J. S. Lim, and D. S\'anchez, Phys. Rev. Lett. \textbf{108}, 246603 (2012).
\bibitem{uts14}
Y. Utsumi, O. Entin-Wohlman, A. Aharony, T. Kubo, and Y. Tokura,
Phys. Rev. B \textbf{89}, 205314 (2014).


\end{thebibliography}
\end{document}